\documentclass[12pt]{article}
\usepackage{amsmath, amssymb, slashed, epsf, color, graphicx}

\begin{document}
\begin{titlepage}
\begin{center}

\hfill IPMU-14-0070 \\

\vspace{2.0cm}
{\large\bf High-Scale SUSY Breaking Models \\
in light of the BICEP2 Result}

\vspace{2.0cm}
{\bf Keisuke Harigaya}$^{(a)}$,
{\bf Masahiro Ibe}$^{(a, b)}$,
{\bf Koji Ichikawa}$^{(a)}$, \\
{\bf Kunio Kaneta}$^{(a)}$,
and
{\bf Shigeki Matsumoto}$^{(a)}$

\vspace{1.5cm}
{\it
$^{(a)}${Kavli IPMU (WPI), University of Tokyo, Kashiwa, Chiba 277-8583, Japan} \\
$^{(b)}${ICRR, University of Tokyo, Kashiwa, Chiba 277-8582, Japan}
}

\vspace{2.5cm}
\abstract{The large value of the tensor-to-scalar ratio in the cosmic microwave background radiation reported by the BICEP2 collaboration gives strong impact on models of supersymmetry (SUSY). The large ratio indicates inflation with a high-energy scale and thus a high reheating temperature in general, and various SUSY models suffer from the serious gravitino and Polonyi problems. In this article, we discuss a class of the high-scale SUSY breaking models which are completely free from those problems. With especially focusing on the dark matter relic abundance, we examine how the BICEP2 result narrows down the parameter space of the models, assuming the simplest chaotic inflation model. We find that the mass of the dark matter is predicted to be less than about 1\,TeV thanks to the non-thermal production in the early universe through the decay of abundant gravitinos produced after the reheating process. We also discuss implications in some details to dark matter searches at collider and indirect dark matter detection experiments.}

\end{center}
\end{titlepage}
\setcounter{footnote}{0}

\subsection*{Introdcution}

The Higgs boson mass of about 126\,GeV observed at the LHC experiment~\cite{higgs mass ex} is rather heavier than the prediction of the conventionally studied minimal supersymmetric standard model (MSSM) with the superparticle masses of ${\cal O}$(0.1--1)\,TeV. This rather large Higgs boson mass and the so far null-observation of the superparticles at the experiment seem to suggest models with rather heavy sparticles of ${\cal O}$(10--100)\,TeV, i.e. the high-scale supersymmetry (SUSY) breaking models~\cite{higgs mass th, Okada:1990gg}. The models fitting the observed Higgs boson mass well are classified into the following two: {\bf (1)} the models in which the masses of the scalars and the gauginos in the MSSM are close with each other, i.e. $m_{\rm gaugino} \simeq m_{\rm scalar} = {\cal O}$(10--100)\,TeV, and {\bf (2)} those in which the gaugino masses are rather suppressed compared to the scalars, i.e. $m_{\rm gaugino} \ll m_{\rm scalar} = {\cal O}$(10--100)\,TeV~\cite{Wells:2004di}-\cite{Arvanitaki:2012ps}.\footnote{When $\tan\beta$, which is the ratio of the vacuum expectation values between up- and down-type Higgs doublets, is extremely close to 1, the observed Higgs boson mass allows us to have models with much higher SUSY breaking scales. See, for examples, references~\cite{Fox:2002bu}-\cite{Ibe:2013rpa}.} These models are particularly favored combined with the cosmological gravitino problem~\cite{gravitino problem}. That is, when the scalar masses are expected to be as large as the gravitino mass as in gravity mediation, the large scalar masses amount to a heavy gravitino, which decays well before the Big-Bang Nucleosynthesis starts~\cite{kkm}.\\

\noindent
Recently, the BICEP2 collaboration reported the first measurement of the tensor-to-scalar ratio of $r = 0.20^{+0.07}_{-0.05}$ in the cosmic microwave background (CMB) radiation~\cite{Ade:2014xna}. Such a large ratio or a corresponding very large Hubble parameter during inflation, $H_{\rm inf} \sim 10^{14}$\,GeV, gives strong impact on the SUSY models. In fact, as discussed in reference~\cite{Ibe:2006am}, such a high inflation scale is quite incompatible with the models with $m_{\rm gaugino} \simeq m_{\rm scalar}$ like the conventional gravity mediation models. For models with $m_{\rm gaugino} \simeq m_{\rm scalar}$, the SUSY breaking field $Z$ is required to be neutral under any symmetry, and hence it is expected to have a large linear term in the K\"ahler potential, $K \simeq c Z + h.c.$, with $c$ being of order the Planck scale. In the presence of such a linear term, the scalar potential of the SUSY breaking field obtains a large linear term during inflation, with which the SUSY breaking field gets shifted to a large expectation value. As a result, the models with $m_{\rm gaugino} \sim m_{\rm scalar}$ suffer from a serious entropy problem, i.e. the Polonyi problem~\cite{Coughlan:1983ci}.\footnote{It is worth notifying that this Polonyi problem cannot be solved by making the mass of the Polonyi field large by some interactions, because the linear term during inflation is very huge, and hence the SUSY breaking field $Z$ is shifted to the Planck scale during inflation at which the interactions responsible for the mass of the Polonyi field are ineffective~\cite{Ibe:2006am}.}\\

\noindent
Interestingly, the second class of the high-scale SUSY breaking models are free from the Polonyi problem, since it can be constructed without having a singlet SUSY breaking field~\cite{Giudice:1998xp}. The models are therefore consistent with the measured tensor-to-scalar ratio. In such models, the scalar bosons obtain SUSY breaking masses from the SUSY breaking sector via tree-level interactions of supergravity, and they are expected to be of the order of the gravitino mass. The gaugino masses are, on the other hand, generated at one-loop level mainly from so-called the anomaly mediated SUSY breaking (AMSB) contributions~\cite{Giudice:1998xp, Randall:1998uk}.\\

\noindent
In this article, we investigate the implications of the observed tensor-to-scalar ratio to the high-scale SUSY breaking models of the second class in more details. In our discussion, we take the simplest realization of the chaotic inflation model~\cite{Linde:1983gd} in supergravity~\cite{Kawasaki:2000yn, Kallosh:2011qk} as an example. We also focus on the models with the simplest origin of the $\mu$-term of the Higgs doublets~\cite{PGM}, where it is generated by tree level interactions to the order parameter of the $R$-symmetry breaking~\cite{mu-term}, and hence it is of the order of the gravitino mass.\footnote{These models are now dubbed the pure gravity mediation (PGM)~\cite{PGM} or the minimal split SUSY models~\cite{minimal split}. In what follows, the high-scale SUSY breaking models refer to these models.} As we will show, the reheating temperature in this class of models is rather high, which is favorable for successful leptogenesis scenario~\cite{leptogenesis}. We will show that the neutralino dark matter density obtains sizable non-thermal contributions from the decay of the gravitino which is abundant for the high reheating temperature. As a result, we find that the mass of the neutralino dark matter is less than 1\,TeV. We also discuss implications in some details to dark matter searches at collider and indirect dark matter detection experiments.\\

\noindent
The article is organized as follows. In next section, we first briefly review the simplest realization of the chaotic inflation with a quadratic potential in supergravity, and discuss the most relevant operators for the decay of the inflaton 
in the high-scale SUSY breaking models (the pure gravity mediation (PGM) and the minimal Split SUSY models). We next discuss details of the neutralino dark matter paying particular attention to the non-thermal contributions to the dark matter density from the gravitino decay. The last section is devoted to conclusions.

\subsection*{Chaotic inflation in supergravity}

Recently, the BICEP2 collaboration has reported an observation of the large scale B-mode in the CMB, which is consistent with a large tensor-to-scalar ratio of $r = {\cal O}(0.1)$ in the cosmic perturbation~\cite{Ade:2014xna}. This large tensor-to-scalar ratio indicates a large inflation scale, which strongly suggests so-called the chaotic inflation model~\cite{Linde:1983gd}. In this section, we review the simplest chaotic inflation model in supergravity and discuss the reheating temperature of the universe expected in this model.\\

\noindent
In order to make our discussion concrete, we assume the simplest chaotic inflation model with a quadratic potential of the inflaton $\phi$, $V = m^2 \phi^2 /2$, where $m$ denotes the mass of the inflaton. In supergravity, this form of the potential is realized by introducing two chiral multiplets $X$ and $\Phi$ that have the following K\"ahler potential and the superpotential~\cite{Kawasaki:2000yn},
\begin{eqnarray}
K = K\left( X X^\dagger, (\Phi + \Phi^\dagger)^2 \right)
= X X^\dagger + \frac{1}{2} (\Phi + \Phi^\dagger)^2 + \cdots,
\qquad
W = m X \Phi,
\label{inflaton potential}
\end{eqnarray}
where `$\cdots$' denotes terms which are irrelevant for our discussion. Note here that the K\"ahler potential possesses a shift symmetry: $\Phi$ transforms as $\Phi \rightarrow \Phi + i \alpha$ under the symmetry with $\alpha$ being a real number. This symmetry plays a crucial role to realize the slow-roll inflation in the chaotic inflation scenario where the field value of the inflaton excesses the Planck scale. The inflaton filed $\phi$ is then identified with the imaginary part of the scalar component of $\Phi$. The shift symmetry is explicitly broken by the spurion field $m$, which transforms as $m \rightarrow m \Phi / (\Phi + i \alpha)$ under the symmetry. With the explicit breaking, the inflaton obtains the quadratic potential. In addition to the approximate shift symmetry, we also impose discrete symmetries, under which $\Phi$ and $X$ transform nontrivially. The discrete symmetries are mandatory to avoid the over-production of the gravitinos at the decay of the inflaton~\cite{over-production}. As we will discuss below, the choice of the discrete symmetries is important to determine the reheating temperature.\\

\noindent
The power spectrum ${\cal P}_{\zeta}$ of the curvature perturbations $\zeta$ is given by~\cite{Lyth:1998xn}, ${\cal P}_{\zeta} = (m^2N_e)/(6\pi^2M_{\rm PL}^2)$, where $N_e$ is the number of e-folding and $M_{\rm PL} \simeq 2.4 \times 10^{18}$\,GeV the Planck scale. From the observed power spectrum $\ln (10^{10}{\cal P}_{\zeta}) \simeq 3.1$~\cite{Ade:2013uln}, we find that the mass of the inflaton is fixed to be $m \simeq 1.6 \times 10^{13}$\,GeV.\footnote{In the chaotic inflation model with a dynamically generated fractional power potential, the inflaton mass can be much higher~\cite{Harigaya:2012pg}, which leads to a higher reheating temperature.} It should be noted that the tensor-to-scalar ratio predicted in this model, $r = 8/N_e$, is consistent with the BICEP2 result for $N_e \simeq 50-60$.\\

\noindent
Let us now discuss the decay of the inflaton. As long as $\Phi$ and $X$ are only the fields having a non-vanishing charge of the discrete symmetry, the inflaton never decays and the universe is not reheated. We therefore consider the cases that some of the MSSM fields are also charged under the discrete symmetries in the inflation sector, so that the inflaton decays into the MSSM fields.\footnote{The decay by the explicit breaking of the discrete symmetry is considered in reference~\cite{Kawasaki:2000ws}.} We also assume that interactions between the inflaton sector and the MSSM fields are controlled by ${\cal O}(1)$ coefficients and the Planck scale if they have mass dimensions. With these assumptions, we classify operators dictating the inflaton decay by their mass-dimensions. When the decay is induced by a dimension-$n$ Planck-suppressed operator ($n \geq 4$), the reheating temperature $T_R$ is estimated to be
\begin{eqnarray}
T_R \simeq 0.26 \, \sqrt{\Gamma_\phi M_{\rm PL}} \simeq
\left\{
\begin{array}{ll}
3.2\,c\,\times 10^{14}\,{\rm (GeV)} & (n = 4)\\
2.1\,c\,\times 10^{9}\,{\rm (GeV)} & (n = 5)\\
1.4\,c\,\times 10^{4}\,{\rm (GeV)} & (n = 6)\\
\end{array}
\right. .
\label{eq: TR}
\end{eqnarray}
Here, the decay width of the inflaton is estimated as $\Gamma_\phi \simeq (c^2\,m/8\pi)\,(m/M_{\rm PL})^{2n-8}$ with $c$ being an ${\cal O}(1)$ coupling constant, assuming two-body decays.\footnote{For the case of $n=4$, $T_R$ is estimated to be higher than the mass of the inflaton. In such a case, interactions between decay products and the inflaton affect the reheating process, and $T_R$ is modified from the above estimation~\cite{Yokoyama:2004pf}-\cite{Mukaida:2012qn}.
It turns out that $T_R$ is still at least as large as the mass of the inflaton, even if we take the effect of the interactions into account.}\\

\noindent
For the case of $n = 4$, the reheating temperature is so large that the universe is over-closed by the non-thermal dark matter production from the gravitino decay, as can be seen in equation~(\ref{eq: NT DM production}). On the other hand, for the case of $n \geq 6$, the reheating temperature is so small that a scenario to generate the baryon asymmetry of the universe is limited. For $n=5$, on the other hand, the reheating temperature is consistent with the one required for successful thermal leptogenesis, namely $T_R \geq 2 \times 10^9$\,GeV~\cite{leptogenesis, Buchmuller:2004nz}. With this success, we focus on the case for $n=5$ in the following arguments. Furthermore, as we will see in next section, the dark matter mass is predicted to be ${\cal O}(1)$\,TeV to be consistent with the observed dark matter relic abundance thanks to the non-thermal production form the gravitino decay.\\

\begin{table}[t]
\begin{center}
\begin{tabular}{cc|ccccccccccc}
& & $\Phi$ & $X$ & $m$ & $H_u$ & $H_d$ & $Q$ & $\bar{u}$ & $\bar{d}$ & $L$ & $\bar{e}$ & $N$ \\
\hline
Case 1 & $Z_{4}$ & $0$ & $3$ & $1$ & $2$ & $2$ & $1$ & $1$ & $1$ & $1$ & $1$ & $1$ \\
&$Z_{4R}$ & $2$ & $0$ & $0$ & $2$ & $2$ & $2$ & $2$ & $2$ & $2$ & $2$ & $2$ \\
\hline
Case 2 & $Z_{4}$ & $2$ & $1$ & $1$ & $0$ & $0$ & $2$ & $2$ & $2$ & $2$ & $2$ & $2$ \\
&$Z_{4R}$ & $0$ & $2$ & $0$ & $2$ & $2$ & $2$ & $2$ & $2$ & $2$ & $2$ & $2$ \\
\hline
\end{tabular}
\end{center}
\caption{\sl \small The charge assignments of the MSSM fields ($H_u$, $H_d$, $Q$, $\bar{u}$, $\bar{d}$, $L$, $\bar{e}$), the right-handed neutrino fields ($N$), and the fields in the inflaton sector ($\Phi$, $X$).}
\label{tab: charge}
\end{table}

\noindent
Now let us show some models where the inflaton decays via dimension $5$ operators. For example, let us consider the discrete symmetries in the inflaton sector given in Table~\ref{tab: charge}, where the charge assignments of the MSSM fields, the right-handed neutrinos, and the fields in the inflaton sector are shown. In Case 1, the $Z_4$ symmetry is a discrete subgroup of the linear combination of $U(1)_Y$ and $U(1)_{B-L}$ (so-called the fiveness), under which the MSSM fields have following charges;
\begin{eqnarray}
(Q, \bar{u}, \bar{e}): 1, \qquad (L,\bar{d}): -3, \qquad
H_u: -2, \qquad H_d:2, \qquad N:5.
\end{eqnarray}
The $Z_4$ symmetry therefore guarantees the stability of the dark matter particle and the proton. In Case 2, the $Z_4$ symmetry contains the $R$-parity. The inflaton decay is then induced by the following dimension-5 operator in each case:\footnote{Daughter particles of the inflaton are now charged under the $SU(2)_L$ symmetry. As is shown in references~\cite{Davidson:2000er, Harigaya:2013vwa}, the daughter particles are then thermalized by the non-abelian gauge interaction instantaneously, and hence the estimation given in equation~(\ref{eq: TR}) is verified.}
\begin{eqnarray}
K &\supset&
\left\{
\begin{array}{ll}
(c/M_{\rm PL}) X^\dag L H_u + {\rm h.c.} & ({\rm Case}~1), \\
(c/M_{\rm PL}) (\Phi + \Phi^\dag) L H_u + {\rm h.c.} & ({\rm Case}~2)\ .
\end{array}
\right.
\label{eq: decay}
\end{eqnarray}
In both cases, the inflaton decays into a pair of the MSSM fields, which leads to the reheating temperature in equation~\,(\ref{eq: TR}) with $n=5$.\\

\noindent
Before closing this section, let us comment on the case where the discrete symmetry is not introduced. In a class of SUSY breaking models, the inflaton decay into gravitinos can be suppressed~\cite{Nakayama:2012hy}. In such a case, we do not have to impose the discrete symmetry, and the inflaton decays via the dimension-5 operator,
\begin{eqnarray}
K \supset (c/M_{\rm PL}) (\Phi + \Phi^\dag) H_u H_d + {\rm h.c.},
\end{eqnarray}
where the $R$ charge of the operator $H_u H_d$ vanishes. The consistency of the operator with the $R$ symmetry is characteristic of the special High-scale SUSY breaking models, namely the pure gravity mediation and the minimal split SUSY models.

\subsection*{Dark matter candidates in gauginos}

\begin{figure}[t]
\begin{center}
\includegraphics[scale=0.76]{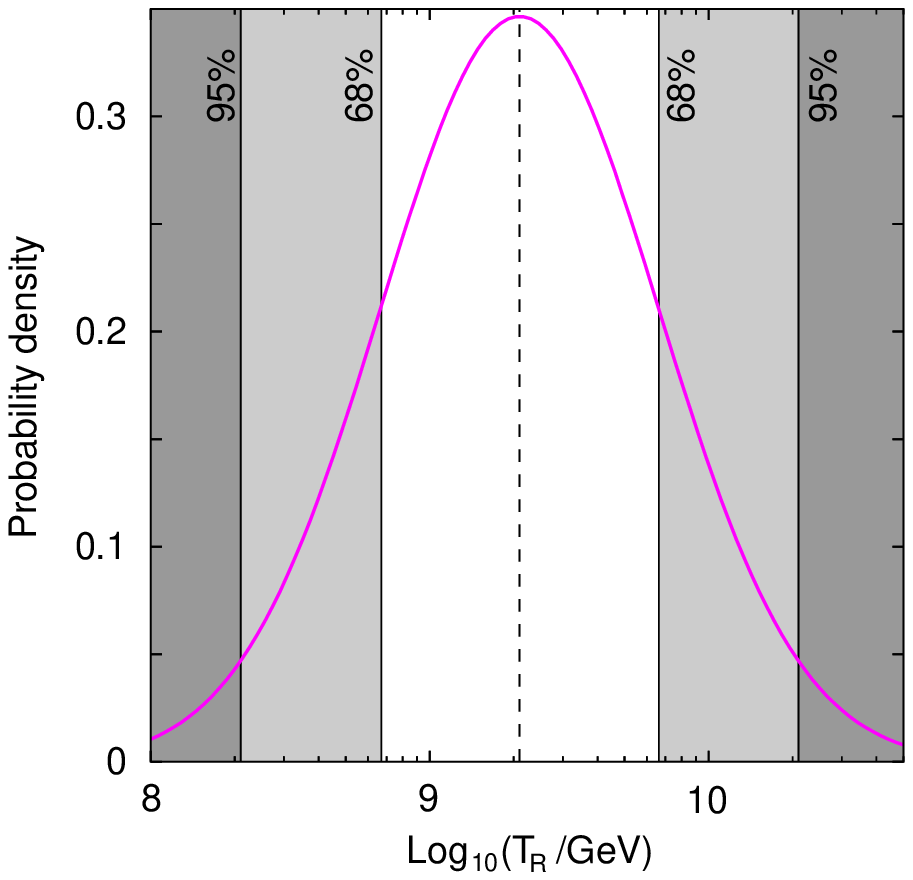}
~~
\includegraphics[scale=0.76]{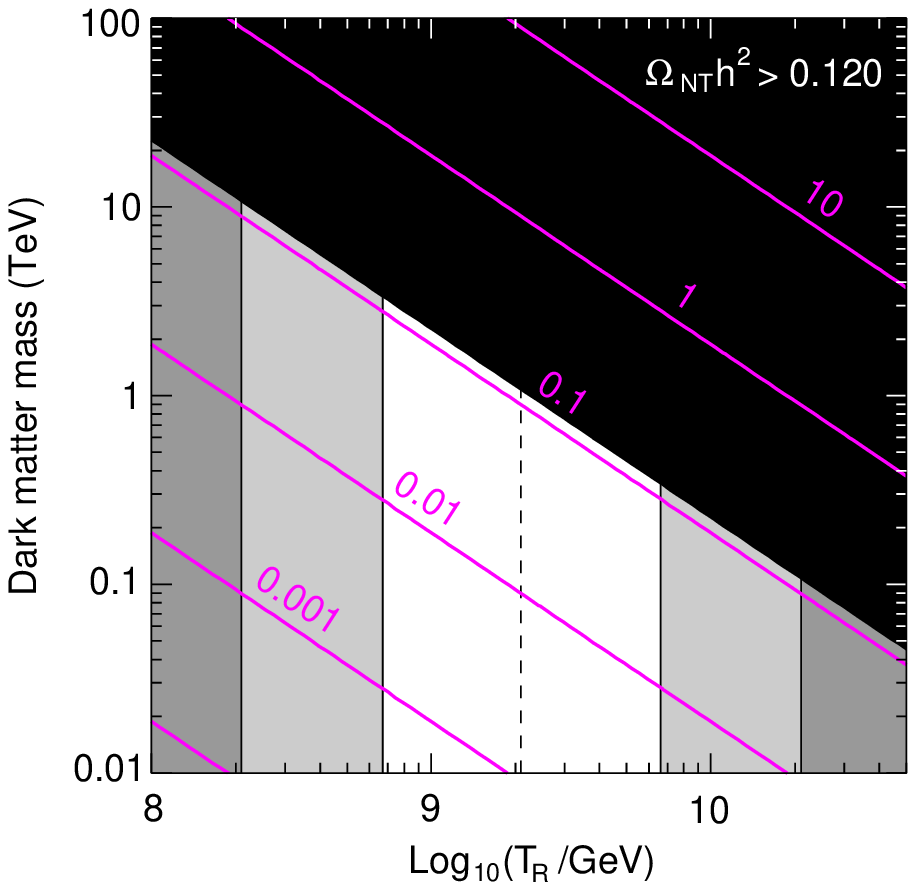}
\caption{\small\sl
{\bf Left panel:} The log-Gaussian probability density of $T_R$ deduced from the BICEP2 result. The light-grey and white regions are the ones favored at 68\% and 95\% confidence level, respectively. See text for more details. {\bf Right panel:} Contours of $\Omega_{\rm NT} h^2$ as a function of $T_R$ and $m_{\rm DM}$. The light-grey and white regions are the same as those in the left panel, while $\Omega_{\rm NT} h^2$ exceeds the observed abundance in the black region.}
\label{fig: probability}
\end{center}
\end{figure}

In the high-scale SUSY breaking models (the PGM type), either the wino or the bino is predicted to be dark matter, and its relic abundance is determined by the sum of two different contributions. One is the contribution from the traditional thermal production ($\Omega_{\rm TH} h^2$)~\cite{thermal wino} and another is the non-thermal production from the gravitino decay ($\Omega_{\rm NT} h^2$)~\cite{non-thermal wino}. The latter contribution depends not only on the dark matter mass ($m_{\rm DM}$) but also the reheating temperature ($T_R$). When the temperature is higher, the more the gravitino is produced, and hence the contribution is larger. The non-thermal contribution is then estimated to be
\begin{eqnarray}
\Omega_{\rm NT} h^2 \simeq
0.16\,\left(m_{\rm DM}/300\,{\rm GeV}\right)\,
\left(T_R/10^{10}\,{\rm GeV}\right).
\label{eq: NT DM production}
\end{eqnarray}
In order to quantitatively discuss how the BICEP2 result affects the PGM type models, we assume that $T_R$ follows the log-Gaussian probability with its mean value and standard deviation being $\ln(2.1 \times 10^9\,{\rm GeV})$ and $(\ln 10)/2$, respectively. The use of such a probability is to take an ${\cal O}(1)$ ambiguity of $T_R$ into account, which is caused by e.g. an ${\cal O}(1)$ change of the coupling constant `c' in equation~(\ref{eq: TR}). We show the probability density $P(\ln T_R)$ in the left panel of Fig.~\ref{fig: probability}, where the probability itself is defined by $P(\ln T_R)\,d(\ln T_R)$. The light-grey and white regions are the ones favored by the BICEP2 result at 68\% and 95\% confidence level, respectively, according to the log-Gaussian probability. The resultant non-thermal contribution $\Omega_{\rm NT} h^2$ is shown in the right panel as a function of $T_R$ and $m_{\rm DM}$. The light-grey and white regions are the same as those in the left panel. It can be seen that, when $T_R \simeq 2.1 \times 10^9\,{\rm GeV}$, the dark matter mass should be less than about 1\,TeV so that $\Omega_{\rm NT} h^2$ does not exceed the observed dark matter density, $\Omega^{\rm (obs)} h^2 \simeq 0.120$~\cite{Ade:2013zuv}.\\

\noindent
The contribution to the dark matter relic abundance from the thermal production does not depend on $T_R$. Instead, it depends on the mass spectrum of the gauginos. In the PGM type models, the gauginos acquire their masses via anomaly mediated contributions and Higgsino threshold corrections~\cite{Giudice:1998xp, Randall:1998uk}. In addition, there are also other contributions when we consider well-motivated extension of the minimal SUSY standard model, i.e. extension with a vector-like matter field and/or a PQ sector~\cite{Nelson:2002sa}-\cite{Harigaya:2013asa}. We therefore treat the gaugino masses as free parameters. Then, remembering the fact that the mixing between the neutral wino and the bino after the electroweak symmetry breaking is much suppressed in the PGM type models, we have the following four possibilities for the spectrum: {\bf (I)} the bino is the lightest SUSY particle (LSP) and the wino is the next-to-lightest SUSY particle (NLSP), {\bf (II)} the bino is the LSP and the gluino is the NLSP, {\bf (III)} the wino is the LSP and the bino is the NLSP,\footnote{Honestly speaking, the charged wino should be called the NLSP, for the neutral wino is the LSP. In order to make our discussion simple, we have used the above designation.} and {\bf (IV)} the wino is the LSP and the gluino is the NLSP. In what follows, we discuss how the BICEP2 result gives impact on these possibilities by calculating the dark matter relic abundance, including the coannihilation between LSP and NLSP particles\footnote{It is also possible to consider the coannihilation in which all the gauginos (the bino, the wino, and the gluino) participate. Since the effect of the BICEP2 result on this possibility can be easily read off from those on other possibilities, we omit to discuss it in this article. Here, it is also worth notifying that the chemical equilibrium between LSP and NLSP is always maintained thanks to decay, inverse-decay, and conversion processes mediated by the Higgsino, the squarks, and the sleptons, even if these particles are as heavy as ${\cal O}(100)$\,TeV.} and the Sommerfeld effect~\cite{sommerfeld effect}-\cite{Harigaya:2014dwa}. Based on obtained results, we also discuss some implications to gaugino searches at collider and indirect dark matter detection experiments.\\

\begin{figure}[t]
\begin{center}
\includegraphics[scale=0.76]{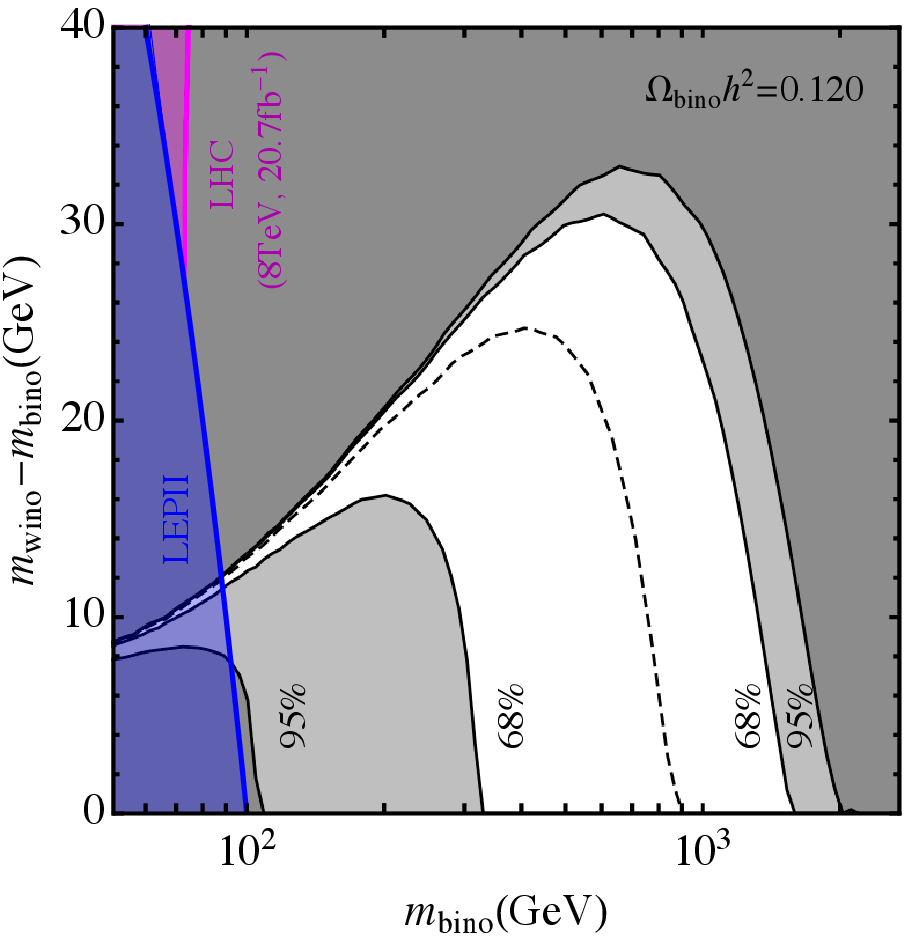}
~~
\includegraphics[scale=0.79]{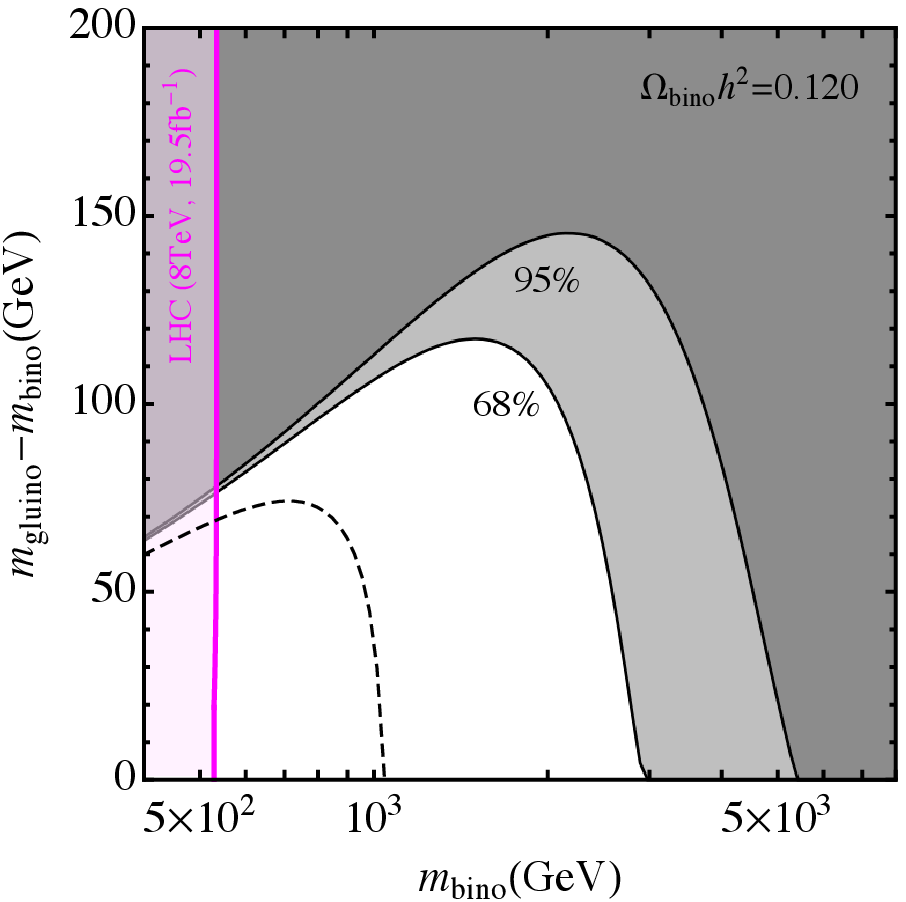}
\caption{\small\sl
{\bf Left panel:} The parameter region favored from the viewpoint of the dark matter relic abundance when the bino is the LSP and the wino is the NLSP. The light-grey and white regions are the same as those in Fig.~\ref{fig: probability}. Limits from the LEP\,II and the LHC experiments at 95\% C.L. are also shown. {\bf Right panel:} The parameter region favored from the viewpoint of the dark matter relic abundance when the bino is the LSP and the gluino is the NLSP. The light-grey and white regions are the same as those in the left panel. A limit from the LHC experiment at 95\% C.L. is also shown. }
\label{fig: bino dark matter}
\end{center}
\end{figure}

\noindent
First, let us consider the case {\bf (I)} where the bino is the LSP and the wino is the NLSP. The parameter region favored from the viewpoint of the dark matter relic abundance is shown in the left panel of Fig.~\ref{fig: bino dark matter} as a function of the bino mass ($m_{\rm bino}$) and the mass difference between the wino and the bino ($m_{\rm wino} - m_{\rm bino}$).
The light-gray and white regions are the same as those in Fig.~\ref{fig: probability}, where the white (light-gray) region is favored by the BICEP2 result at 68\% (95\%) confidence level (C.L.) with assuming the simplest chaotic inflation model. It can be seen that the mass of the bino dark matter is at most around 1\,TeV with the ${\cal O}(10)$\,GeV mass difference thanks to the non-thermal contribution. Some limits obtained by collider experiments are also shown in the panel. The limit painted by blue comes from the LEP\,II experiment, which is obtained by the search for the wino pair production via the radiative return process~\cite{Heister:2002mn}. The limit painted by purple is from the LHC experiment, which is obtained by the search for the production of neutral and charged winos~\cite{Aad:2014nua}. Here we assumed that the process provides three charged leptons with missing energy in its final state at 100\% ratio. This assumption is verified in the most of the parameter region of the PGM type models, especially when the sleptons are somewhat lighter than the Higgsinos and/or the squarks. It is obvious that detection of ${\cal O}(10)$\,GeV soft leptons at 14\,TeV running of the LHC experiment will play an crucial role to detect the bino dark matter.\\

\noindent
Next, we consider the case {\bf (II)} where the bino is the LSP and the gluino is the NLSP. The parameter region favored by the dark matter relic abundance is shown in the right panel of Fig.~\ref{fig: bino dark matter} as a function of the bino mass ($m_{\rm bino}$) and the mass difference between the gluino and the bino ($m_{\rm gluino} - m_{\rm bino}$). The light-gray and white regions are the same as those in the left panel. Thanks to the non-thermal production again, the mass of the dark matter is predicted to be less than about 1\,TeV with the ${\cal O}(100)$\,GeV mass difference when $T_R \simeq 2.1 \times 10^9\,{\rm GeV}$. On the other hand, the dark matter mass is increased to a few TeV when the reheating temperature is somewhat suppressed, because the coannihilation process between gluinos is so efficient. Present limit from the LHC experiment is also shown in the plot as a pink region, which is obtained by the search for the gluino pair production associated with the initial state radiation(s) using 19.5\,fb$^{-1}$ data at 8\,TeV running~\cite{Chatrchyan:2014lfa}. The limit will reach about $m_{\rm gluino} \sim m_{\rm wino} \sim 1$\, TeV in near future using 100\,fb$^{-1}$ data at 14\,TeV running~\cite{Bhattacherjee:2013wna}, so that it can cover almost entire mass region of the bino dark matter when $T_R \simeq 2.1 \times 10^9$\,GeV.\\

\begin{figure}[t]
\begin{center}
\includegraphics[scale=0.77]{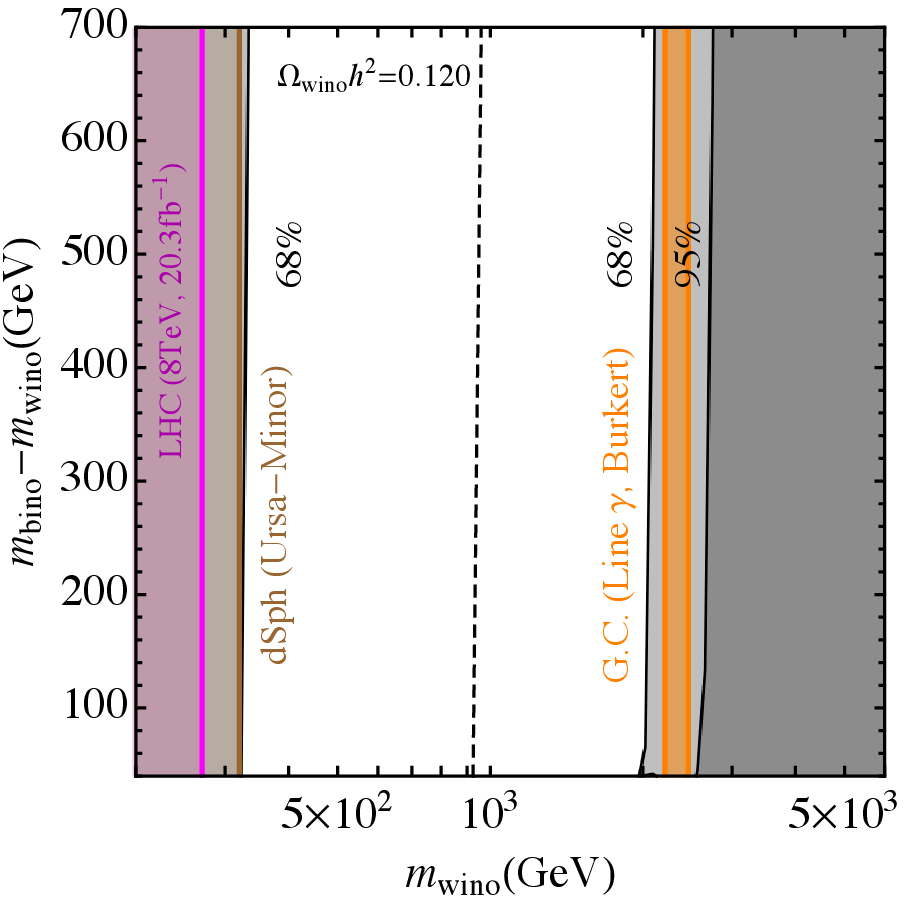}
~~
\includegraphics[scale=0.82]{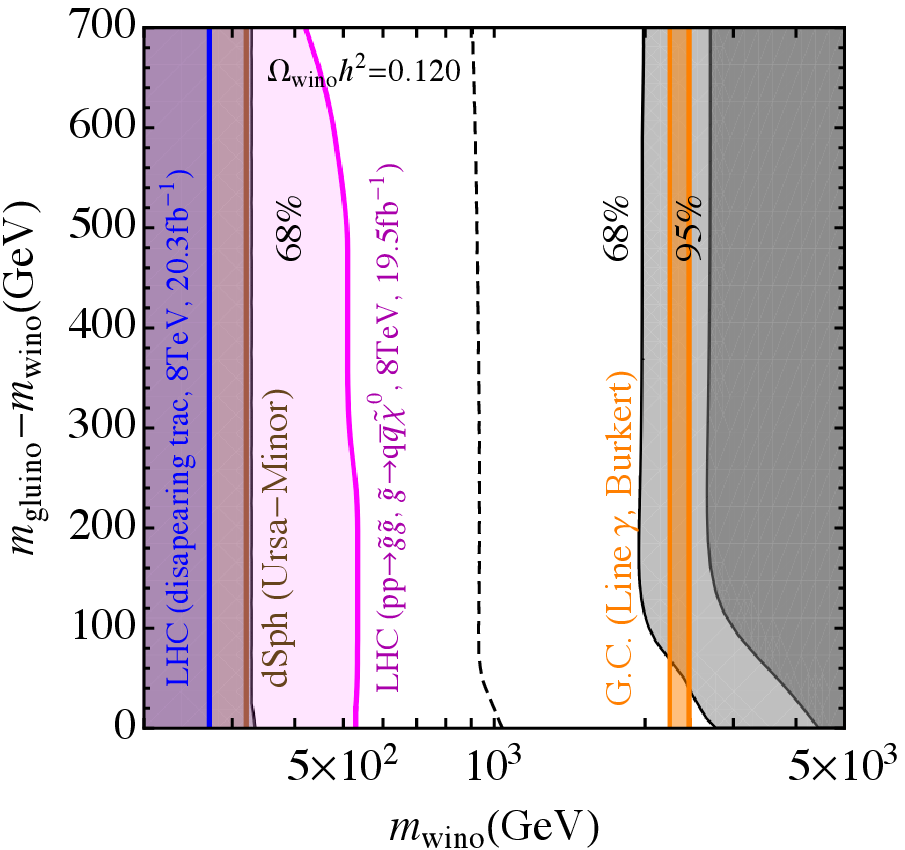}
\caption{\small\sl
{\bf Left panel:} The parameter region favored from the viewpoint of the dark matter relic abundance when the wino is the LSP and the bino is the NLSP. The light-grey and white regions are the same as previous figures. Limits from the LHC and the indirect dark matter detection (two kinds) experiments at 95\% C.L. are also shown. {\bf Right panel:} The parameter region favored from the viewpoint of the dark matter relic abundance when the bino is the LSP and the gluino is the NLSP. The light-grey and white regions are the same as those in the left panel. A limit from the LHC (two kinds) and the indirect dark matter detection (two kinds) experiments at 95\% C.L. are also shown.}
\label{fig: wino dark matter}
\end{center}
\end{figure}

\noindent
The wino is the LSP and the bino is the NLSP in the case {\bf (III)}. The parameter region favored by the dark matter relic abundance is shown in the left panel of Fig.~\ref{fig: wino dark matter} as a function of the wino mass ($m_{\rm wino}$) and the mass difference between the bino and the wino ($m_{\rm bino} - m_{\rm wino}$). The light-gray and white regions are the same as those in previous figures. The dependence of the mass difference on the parameter region is very weak, because the dark matter relic abundance is almost governed by the wino self-annihilation and the non-thermal production. As in the case of the bino dark matter, the mass of the wino dark matter is predicted to be about 1\,TeV when $T_R \simeq 2.1 \times 10^9\,{\rm GeV}$. When the reheating temperature is somewhat suppressed, the predicted mass is shifted to a few TeV, which is the value obtained when the abundance is determined only by the wino thermal production~\cite{thermal wino}.\\

\noindent
Limits from the LHC and the indirect dark matter detection at the Fermi-LAT and the H.E.S.S. experiments are also shown as pink and orange regions, respectively. The LHC limit is from the search for the disappearing charged track caused by the charged wino decay inside inner detectors~\cite{Aad:2013yna}.\footnote{The charged wino is highly degenerate with the neutral wino, so that it decays into a neutral wino by emitting a soft pion with its decay length of about 7\,cm. The mass difference between charged and neutral winos has already been calculated at full two-loop level~\cite{Ibe:2012sx}.} The wino mass up to 500\,GeV will be covered in near future by the search with 100\,fb$^{-1}$ data at 14\,TeV running~\cite{LHCProspect}. On the other hand, the limit from the indirect dark matter detection is obtained by observing gamma-rays from the galactic center (G.C.)~\cite{Abramowski:2013ax} and dwarf spheroidal galaxies (dSphs)~\cite{Ackermann:2013yva}.\footnote{The indirect detection utilizing cosmic-ray anti-protons is potentially important, as clearly pointed out in reference~\cite{Hryczuk:2014hpa}, when systematic errors associated with the use of the diffusion equation are accurately evaluated. The limit on $m_{\tilde{w}}$ could be as strong as $m_{\tilde{w}} > 500$\,GeV.} For the dSph observation, we use only a classical dSph (Ursa-Minor) to put a robust limit. The limit is not altered even if we include other classical dSphs. On the contrary, if we include ultra-faint dSphs in our analysis, the region $m_{\rm wino} < 0.4\,{\rm TeV}$ and $2.13\,{\rm TeV} < m_{\rm wino} < 2.53\,{\rm TeV}$ is excluded. We omit to depict this limit in the plot, because the error of the dark matter profile inside each ultra-faint dSph, namely the error of so called the $J$-factor, is still very large without assuming some relations between its profile and kinematic data~\cite{J-factor for UF dSphs}. In future, the parameter region $m_{\rm wino} < 0.77\,{\rm TeV}$ and $1.91\,{\rm TeV} < m < 2.67\,{\rm TeV}$ ($m_{\rm wino} < 0.84\,{\rm TeV}$ and $1.85\,{\rm TeV} < m_{\rm wino} < 2.7\,{\rm TeV}$) will be covered by the Fermi-LAT experiment after 10 years (15 years) data taking. Here, we assumed that the error of the $J$-factor is reduced to the level of the classical ones in future, namely $\Delta \log_{10}[J(0.5^\circ)/({\rm GeV}^2 {\rm cm}^{-5} {\rm sr})] = 0.20$. In the G.C. observation, it is well known that the signal flux from the dark matter annihilation suffers from large systematic uncertainties due to our limited knowledge of the dark matter profile at the G.C. region~\cite{Nesti:2013uwa}. We thus used the Burket (cored) profile~\cite{Burkert:1995yz} instead of the NFW (cuspy) profile~\cite{Navarro:1995iw} in our analysis in order to put a robust limit.\\

\noindent
Finally, we consider the case {\bf (IV)} where the wino is the LSP and the gluino is the NLSP. The parameter region favored by the dark matter relic abundance is shown in the right panel of Fig.~\ref{fig: wino dark matter} as a function of the wino mass ($m_{\rm wino}$) and the mass difference between the gluino and the wino ($m_{\rm gluino} - m_{\rm wino}$). The light-gray and white regions are the same as those in the left panel. The resultant region is almost the same as that for the case (III) except for the small one in which the wino is degenerate with the gluino and thus the gluino coannihilation process is efficient. Two limits from the LHC experiment are shown as blue and pink regions. The first one (which is the same as that in the left panel of Fig.~\ref{fig: wino dark matter}) comes from the disappearing charged track search~\cite{Aad:2013yna}, while another one (which is the same as that in the right panel of Fig.~\ref{fig: bino dark matter}) is from the search for the gluino pair production~\cite{Chatrchyan:2014lfa}. These two searches will have complimentary roles at 14\,TeV running of the LHC experiment: the latter search will play an crucial role to explore the wino dark matter when the mass difference between the gluino and the wino is large enough, while the former one will be important when the mass difference is suppressed. The limit from the indirect dark matter detection at the Fermi-LAT and the H.E.S.S. experiments are shown as orange lines, which are the same as those in the left panel.\\

\subsection*{Conclusions}

To conclude, the BICEP2 result has given strong impact on SUSY models, because the tensor-to-scalar ratio of $r \simeq 0.2$ means inflation occurred at some high scale. This fact indicates that the reheating temperature is also expected to be high in general, and we have to seriously consider the gravitino and the Polonyi problems in various SUSY models. The high-scale SUSY breaking models (the PGM type) are known to be completely free from the problems, so that the models become more and more attractive than before. In this article, we have therefore discussed the impact of the BICEP2 result on the models especially focusing on the dark matter relic abundance, adopting the simplest chaotic inflation model. Thanks to the non-thermal dark matter production in the early universe through the decay of the gravitinos produced just after the reheating process, the dark matter mass is predicted to be less than about 1\,TeV when $T_R \sim 2 \times 10^9$\,GeV. This result encourages us very much, for the dark matter seems to be detected at collider or dark matter indirect detection experiments in near future.

\subsection*{Acknowledgments}

This work is supported by Grant-in-Aid for Scientific Research from the Ministry of Education, Science, Sports and Culture (MEXT), Japan, No. 24740151 (M.I.), Nos. 23740169 \& 22244021 (S.M.), and also by World Premier International Research Center Initiative (WPI Initiative), MEXT, Japan. The work of K.H. and K.K. is supported by JSPS Research Fellowship for Young Scientists.


\end{document}